\documentclass[fleqn,usenatbib]{mnras}

\usepackage{newtxtext,newtxmath}

\usepackage[T1]{fontenc}

\DeclareRobustCommand{\VAN}[3]{#2}
\let\VANthebibliography\thebibliography
\def\thebibliography{\DeclareRobustCommand{\VAN}[3]{##3}\VANthebibliography}

\usepackage{graphicx}	
\usepackage{amsmath}	
\usepackage{longtable}
\usepackage{array}
\usepackage{threeparttablex} 
\usepackage{graphicx}
\usepackage{subcaption}
\usepackage{amsmath}
\usepackage{xcolor}

\title[Late-Time Light Curves of SN Ia]{The Late-Time Light Curves of Type Ia Supernovae: Confronting Models with Observations}

\author[V. Tiwari et al.]{
Vishal Tiwari,$^{1,2}$\thanks{E-mail: vtiwari@gatech.edu (VT)}
Or Graur,$^{3,4}$
Robert Fisher,$^{1}$
Ivo Seitenzahl,$^{5}$
Shing-Chi Leung,$^{6}$
Ken’ichi Nomoto,$^{7}$
\newauthor{Hagai Binyamin Perets$^{8}$ and Ken Shen$^{9}$}
\\
$^{1}$Department of Physics, University of Massachusetts Dartmouth, 285 Old Westport Road, North Dartmouth, MA 02740, USA\\
$^{2}$Center for Relativistic Astrophysics, School of Physics, Georgia Institute of Technology, Atlanta, GA 30332, USA\\
$^{3}$Institute of Cosmology and Gravitation, University of Portsmouth, Portsmouth, PO1 3FX, UK\\
$^{4}$Department of Astrophysics, American Museum of Natural History, Central Park West and 79th Street, New York, NY 10024, USA\\
$^{5}$School of Science, University of New South Wales, Australian Defence Force Academy, Canberra, ACT 2600, Australia\\
$^{6}$TAPIR, Mailcode 350-17, California Institute of Technology, Pasadena, CA 91125, USA\\
$^{7}$Kavli Institute for the Physics and Mathematics of the Universe (WPI), The University of Tokyo Institutes for Advanced Study, The University of Tokyo,\\ Kashiwa, Chiba 277-8583, Japan\\
$^{8}$Technion - Israel Institute of Technology, Physics Department, Haifa, Israel 32000\\
$^{9}$Astronomy  Department,  University  of  California  Berkeley, Berkeley, CA 94720, USA\\
}

\date{Accepted XXX. Received YYY; in original form ZZZ}

\pubyear{2021}

\begin{document}
\label{firstpage}
\pagerange{\pageref{firstpage}--\pageref{lastpage}}
\maketitle

\begin{abstract}
Type Ia supernovae (SNe Ia) play a crucial role as standardizable candles in measurements of the Hubble constant and dark energy. Increasing evidence points towards multiple possible explosion channels as the origin of normal SNe Ia, with possible systematic effects on the determination of cosmological parameters. We present, for the first time, a comprehensive comparison of  publicly-available SN Ia model nucleosynthetic data with observations of late-time light curve observations of SN Ia events. These models span a wide range of white dwarf (WD) progenitor masses, metallicities, explosion channels, and numerical methodologies. We focus on the influence of $^{57}$Ni and its isobaric decay product $^{57}$Co in powering the late-time ($t > 1000$ d) light curves of SNe Ia. $^{57}$Ni and $^{57}$Co are neutron-rich relative to the more abundant radioisotope $^{56}$Ni, and are consequently a sensitive probe of neutronization at the higher densities of near-Chandrashekhar (near-$M_{\rm Ch}$) progenitor WDs.
We demonstrate that observations of one SN Ia event, SN 2015F is only consistent with a sub-$M_{\rm Ch}$ WD progenitor. Observations of four other events (SN 2011fe, SN 2012cg, SN 2014J, SN2013aa) are consistent with both near-$M_{\rm Ch}$ and sub-$M_{\rm Ch}$ progenitors. Continued observations of late-time light curves of nearby SNe Ia will provide crucial information on the nature of the SN Ia progenitors.
\end{abstract}





\section{Introduction}

Type Ia supernovae (SNe Ia) are thought to be white dwarf stars (WDs) composed primarily of carbon and oxygen, which undergo explosive nuclear burning.  SNe Ia are important across many astrophysical domains, serving as standardizable candles for cosmology \citep {rust1974use,pskovskii1977light, phillips93},  sources of cosmic rays \citep {baadezwicky34, ginzburg64, drury12}, turbulence \citep {elmegreenscalo04}, enriched isotopes for the interstellar medium \citep{nomoto1984accreting,thielemann1986explosive}, and endpoints of binary evolution.

An isolated WD is inherently stable. Therefore, virtually all SNe Ia explosion models call for a companion star.\footnote {One exception being the pycnonuclear-driven model of \cite {chiosietal15}} However, the nature of the companion star and the explosion mechanism of SNe Ia is still unclear. The most frequently-discussed possibilities for the companion are a main-sequence or a red-giant star in the single-degenerate (SD) channel \citep {whelaniben73}, or another white dwarf in the double-degenerate (DD) channel \citep {webbink84, ibentutukov84, nomoto82}. 

In this paper, we explore how late-time ($t > 1000$ d) observations of the light curves of SNe Ia constrain their stellar progenitors and explosion mechanisms. An important distinction between explosion channels originates from the importance of electron capture reactions in the dense cores of near-Chandrashekhar (near-$M_{\rm Ch}$) WDs originating in the SD channel. The process of neutronization  leads to an overall greater prevalence of neutron-rich isotopes in near-$M_{\rm Ch}$ SD events, including the radioisotopes  $_{28}^{57}\textrm{Ni}$ and $_{27}^{55}\textrm{Co}$ \citep {nomoto1984accreting,thielemann1986explosive, seitenzahletal13}. These neutron-rich radioisotopes decay over much longer timescales than the well-known $^{56}$Ni chain energizing early SNe Ia light curves \citep {pankey62}, and consequently power the SN Ia light curves at late times. The $A = 57$ isobar decay chain has a half-life of just over nine months:  $_{28}^{57}\textrm{Ni}\;\overset{\mathrm{1.5 d}}\longrightarrow\;_{27}^{57}\textrm{Co}\;\overset{\mathrm{272 d}}\longrightarrow\;_{26}^{57}\textrm{Fe}$. In addition, the $A = 55$ isobar radioisotopes decay to $^{55}$Mn over several years:
 $_{27}^{55}\textrm{Co}\;\overset{\mathrm{18 h}}\longrightarrow\;_{26}^{55}\textrm{Fe}\;\overset{\mathrm{3 y}}\longrightarrow\;_{25}^{55}\textrm{Mn}$.
 In practice, the observational challenge of making accurate late-time light curve measurements of SNe Ia beyond 1000 days limits the use of this method to nearby well-studied SNe Ia followed up with \textit{Hubble Space Telescope} (HST). To date, the late-time light curve technique has been applied to six SN Ia events: SN 2011fe \citep {shappeeetal16,dimitriadis2017late,kerzendorf2017extremely}, SN 2012cg \citep {childressetal15, graur2016late}, SN 2013aa \citep{jacobson2018constraining}, SN 2014J \citep{graur2018late, yang2018mapping, li2019observations}, ASASN-14lp \citep{graur2018late_ASASSN} and SN 2015F \citep{graur2018observations}. Abundance ratios of $^{57}$Ni/$^{56}$Ni have been obtained for five of these (SNe 2011fe, 2012cg, 2013aa, 2014J and 2015F), and ratios of $^{55}$Fe/$^{57}$Co for three (SNe 2011fe, 2013aa, and 2014J).

Comparing model predictions against the abundances inferred from late-time light curves allows us to address key outstanding questions surrounding the nature of SNe Ia.
These questions include: Are the stellar progenitors of SNe Ia sub-$M_{\rm Ch}$ or near-$M_{\rm Ch}$ WDs? What is the explosion mechanism of SNe Ia? In the context of near-$M_{\rm Ch}$ WDs, how do the WDs ignite and subsequently detonate?  In the context of sub-$M_{Ch}$ WDs, do only relatively massive $M \sim$ 1 $M_{\odot}$ CO WDs successfully detonate, or can lower mass WDs also detonate, possibly through thin helium layers?

SNe Ia exhibit a diverse range of luminosities and nucleosynthetic yields, both of the key radioactive isotope $^{56}$Ni which powers early light curves, as well as of the trace radioisotopes $^{55}$Co and $^{57}$Ni which dominate late-time light curves. In this paper, we will explore the full diversity within all major channels of SNe Ia, ranging from sub-Chandrasekhar (sub-$M_{\rm Ch}$) WDs detonating through double detonations and  double-degenerate mergers, through near-$M_{\rm Ch}$ WDs detonating within the SD channel, using a large dataset compiled from the literature (see table \ref{tab:obsTab} for a summary). In several instances, we have included similar physical models computed by different groups using different simulation codes, in order to address the robustness of our conclusions to different numerical methodologies.

While previous work confronted a single supernova with a few  models, this is the first paper to confront the entire sample of late-time SN Ia observations with a comprehensive range of models. With this work, we address the question of how theoretical models of SN Ia cope with an ensemble that represents a swath of the whole normal SN Ia population.\footnote {In this paper, we broadly consider a wide range of explosion channels and computational methodologies for which detailed nucleosynthetic yields are publicly available. It should be noted that there are some additional explosion channels considered in the literature, most notably the core degenerate channel \citep{kashi2011circumbinary,soker2019supernovae}, for which detailed nucleosynthetic yields have not yet been obtained.} Further, we also address the WD stellar progenitor parameters and explosion channels which can be inferred from late-time observations.

The format of the paper is as follows. In section \ref{sec:57_56_theory}, we discuss how the isobaric ratio of $^{57}$Ni/$^{56}$Ni (57/56) nuclides can be used to distinguish between near-{$M_{\rm{Ch}}$} and sub-{$M_{\rm{Ch}}$} progenitors. We also investigate how key stellar parameters, including the WD stellar progenitor metallicity, progenitor central density, and ignition can affect this 57/56 isobaric ratio. In section \ref{sec:lateTimeSimandObs}, we describe the details of the numerical simulations of the openly available models used and observational events used in this work. We present our results comparing the late-time data against models in section \ref{sec:LateTimeresults}. In section \ref{sec:LateTimeConclusion}, we discuss the conclusions of our findings.

\section{$^{57}$Ni/$^{56}$Ni vs $^{56}$Ni Theory}
\label{sec:57_56_theory}

Constraining the parameters of individual SN Ia from the late-time light curve is not trivial because of the very high-dimensional parameter space spanned by explosion models. We reduce the dimensionality of the model output space to only two parameters: $^{56}$Ni and $^{57}$Ni/$^{56}$Ni, under the assumption that the late-time light curve is mainly powered by the radioactive decay of $^{57}$Ni chain nuclides.

In the following section, we discuss how the ratio $^{57}$Ni/$^{56}$Ni varies within the SN Ia parameter space, including the nature of the WD progenitor (sub-$M_{\rm Ch}$ or near-$M_{\rm Ch}$), the stellar progenitor metallicity, central density and ignition in case of a near-$M_{\rm{Ch}}$ progenitor. We illustrate these effects on $^{57}$Ni/$^{56}$Ni using the near-$M_{\rm Ch}$ models from \citet{daveetal17} (Figure \ref{fig:densityLatetime}) and sub-$M_{\rm Ch}$ models from \citet{shen2018sub}(Figure \ref{fig:subChar}).

\subsection {WD Progenitors: sub-$M_{\rm Ch}$ and near-$M_{\rm Ch}$}
\label{subsec:near_vs_subCh_theory}

\begin{figure*}
	\includegraphics[width=1.05\textwidth]{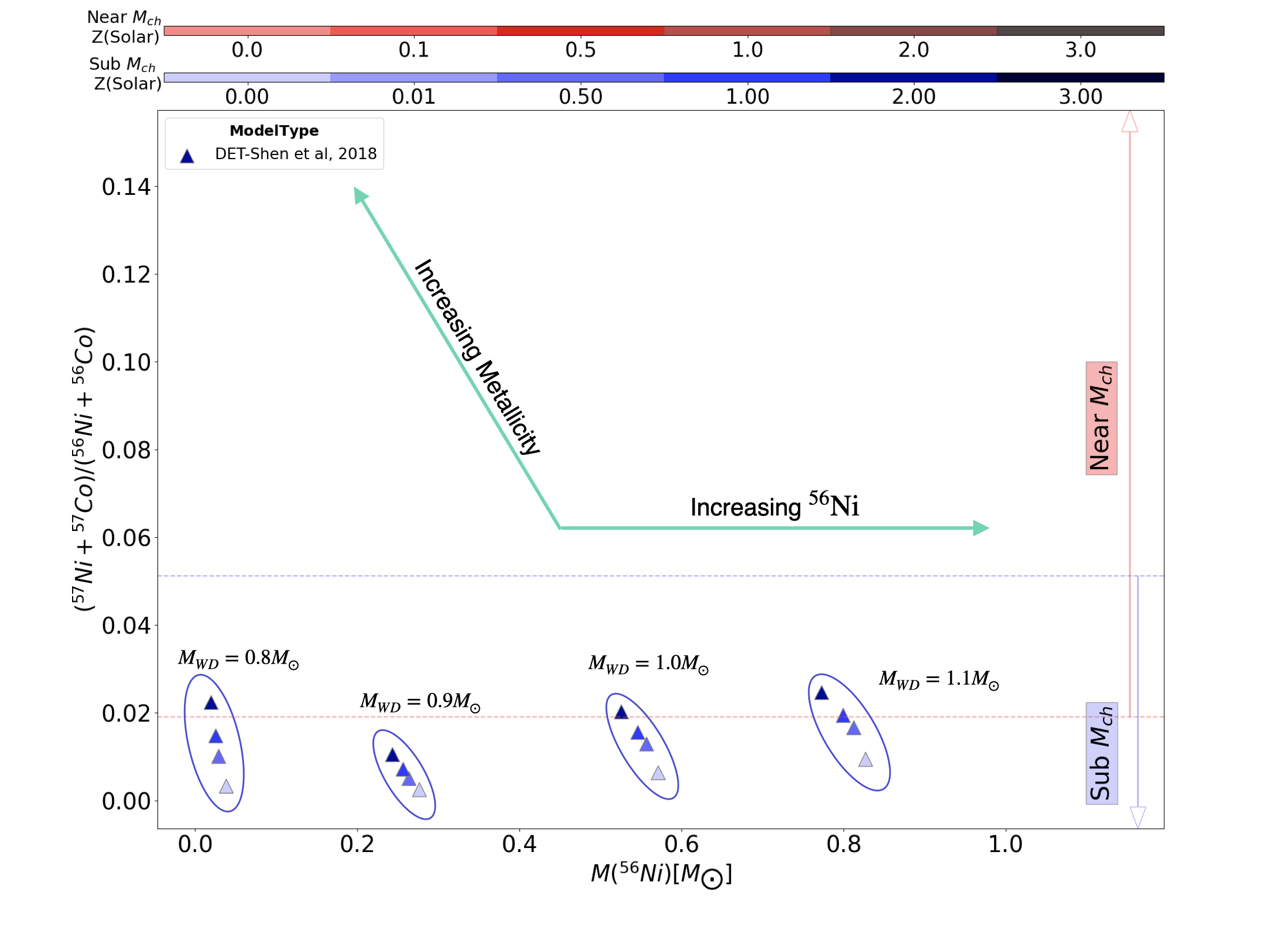}
	\caption{Plot showing the ratio ($\rm{^{57}Ni}$+$\rm{^{57}Co}$)/($\rm{^{56}Ni}$+$\rm{^{56}Co}$) against $M(\rm{^{56}Ni})$ $M_{\odot}$ for the sub-$M_{\rm{Ch}}$ model of \citet{shen2018sub} with varying metallicity and progenitor mass. The shade of blue represents the metallicity, and we see that the ratio 57/56 increases with an increasing metallicity. The horizontal blue line marks the upper bound on 57/56 ratio among all the sub-$M_{\rm{Ch}}$ models, and the red horizontal line shows the lower bound on all the near-$M_{\rm{Ch}}$ models considered in this work. All the solar and sub-solar metallicity models of \citet{shen2018sub} lie below the red-line, the "only sub-$M_{\rm{Ch}}$ region," while some super solar metallicity models are in the "overlap region."
	\label{fig:subChar}}
\end{figure*}

\begin{figure*}
    \includegraphics[width=1.1\textwidth]{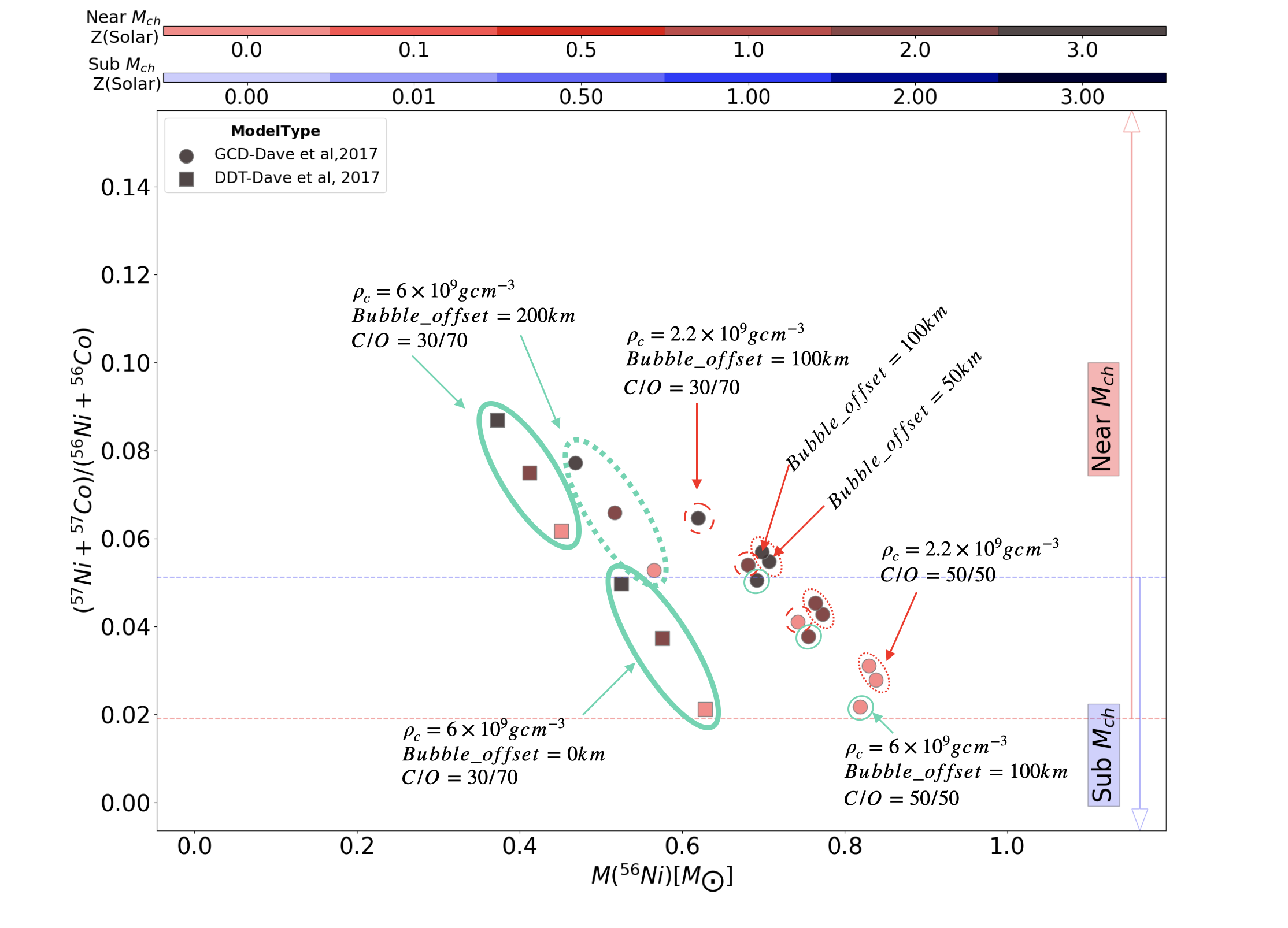}
	\caption{Plot showing the ratio ($\rm{^{57}Ni}$+$\rm{^{57}Co}$)/($\rm{^{56}Ni}$+$\rm{^{56}Co}$) against $M(\rm{^{56}Ni})$ for the near-$M_{\rm{Ch}}$ model of \citet{daveetal17} with varying metallicity (Z), central density ($\rho_{c}$), C/O fraction, bubble offset and bubble radius. The bubble offset and bubble radius are the location of the ignition from the center of the WD and the radius of the ignition bubble respectively. The models in green oval are high-density DDT and GCD models with $\rho_{c}$ = 6 $\times$ 10$^9$ g cm$^{-3}$, while the models in red ovals are lower density models with $\rho_{c}$ = 2.2 $\times$ 10$^9$ g cm$^{-3}$. The DDT models in thick solid green ovals have unequal C/O fraction of 30/70 with centrally ignited detonation (bubble offsets = 0km) and an off-center detonation (bubble offset of 200km). The GCD models in dotted green oval have unequal C/O fraction of 30/70 and an off-center detonation with bubble offset of 200km. The thin solid green ovals are GCD models with an equal fraction of C/O (50/50) and a bubble offset of 100km. The dashed and dotted red models show GCD models with a C/O fraction of 30/70 and 50/50, respectively. As shown, red oval groups have models with 50km and 100km bubble offsets. The horizontal blue line marks the upper bound on 57/56 ratio among all the sub-$M_{\rm{Ch}}$ models, and the red horizontal line shows the lower bound on all the near-$M_{\rm{Ch}}$ models considered in this work. Around half of the models of \citet{daveetal17} lie in the "only near-$M_{\rm{Ch}}$ region," while the other half lies in the "overlap region." \label{fig:densityLatetime} }
\end{figure*}

The electron capture rate is higher and more sensitive in near-$M_{\rm{ch}}$ environments as compared to sub-$M_{\rm{Ch}}$ environments \citep{bravo2019sensitivity}. Thus, we expect the ratio of 57/56 for near-$M_{\rm Ch}$ progenitors to be much higher than for the sub-$M_{\rm{Ch}}$ progenitors. This distinction can be qualitatively seen in Figure \ref{fig:densityLatetime} and Figure \ref{fig:subChar}, where the near-$M_{\rm{Ch}}$ models (shown in red in Fig \ref{fig:densityLatetime}) typically have on average a higher 57/56 ratio as compared to the sub-$M_{\rm{Ch}}$ models (shown in blue in Fig \ref{fig:subChar}). This distinction in the 57/56 ratio shows how the the late-time light curves can be used as a probe to constrain the parameters of the progenitors of SNe Ia.

Central densities also play a crucial role in the production of $\rm{^{55}Co}$. For densities $\rho \gtrsim 2 \times 10^8$ g cm$^{-3}$, yields reach normal NSE freeze-out because of lower entropy, which in turn leads to a higher abundance of $\rm{^{55}Co}$ \citep{thielemann1986explosive, bravo2012sensitivity}. On the other hand, densities $\rho \lesssim  2 \times 10^8$ g cm$^{-3}$ lead to an alpha-rich freeze-out due to higher entropy and $\rm{^{55}Co}$ doesn't survive the freeze-out. Therefore we expect higher fractions of $\rm{^{55}Co}$ (which is parent nuclei of stable monoisotope $\rm{^{55}Mn}$) in the denser near-$M_{\rm{Ch}}$ progenitors as compared to the lighter sub-$M_{\rm{Ch}}$ progenitors (Figure \ref{fig:55by56plot}).


\begin{figure*}
  \centering
  \includegraphics[width=0.48\linewidth]{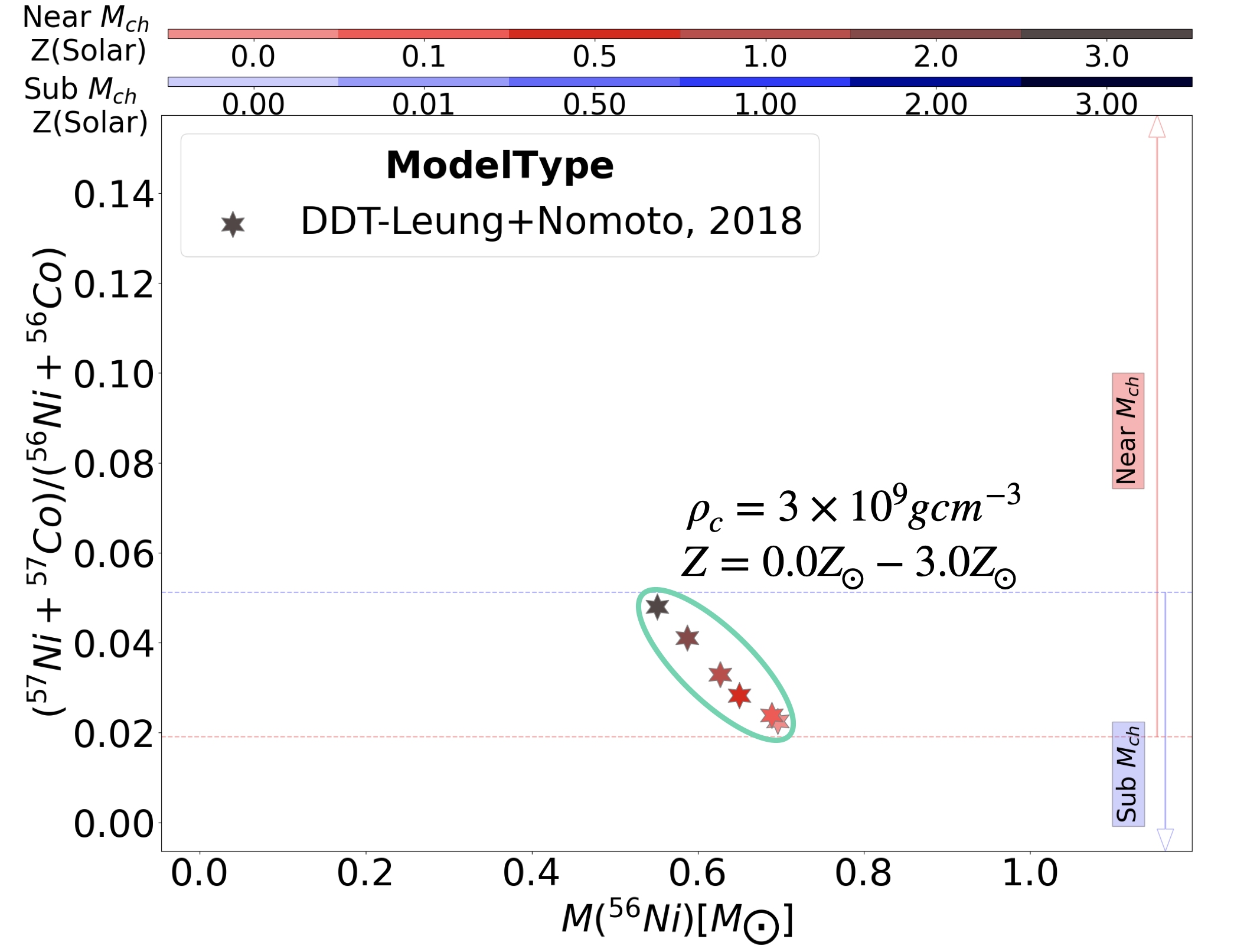} \qquad
  \includegraphics[width=0.48\linewidth]{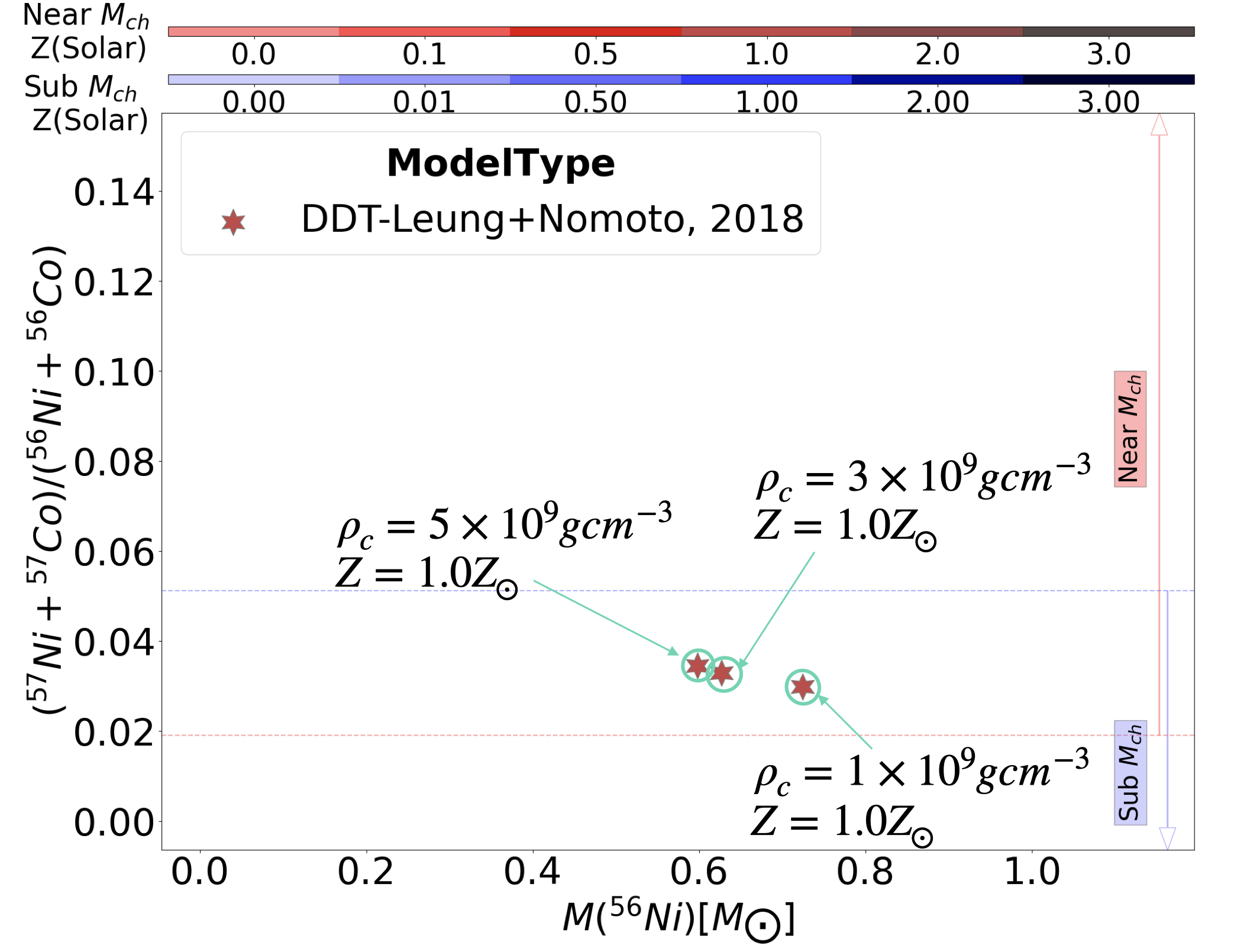}

  \caption{Plots showing the ratio ($\rm{^{57}Ni}$+$\rm{^{57}Co}$)/($\rm{^{56}Ni}$+$\rm{^{56}Co}$) against $M(\rm{^{56}Ni})$ for the near-$M_{\rm{Ch}}$ models of \citet{leung2018explosive}. Plot (a) shows DDT models with central density $\rho_c = 3.0 \times 10^9$ g cm$^{-3}$ and metallicities ranging from Z = 0.0 Z$_{\odot}$ to 3.0 Z$_{\odot}$. Plot (b) shows DDT models with metallicity Z = 1.0 Z$_{\odot}$ and central density ranging from $\rho_c = 1.0 \times 10^{9}$ g cm$^{-3}$ - $5.0 \times 10^{9}$ g cm$^{-3}$. The horizontal blue line marks the upper bound on 57/56 ratio among all the sub-$M_{\rm{Ch}}$ models, and the red horizontal line shows the lower bound on all the near-$M_{\rm{Ch}}$ models considered in this work. All these sets of models lie in the "overlap region".}
  \label{fig:figureTest}
\end{figure*}

\subsection {Stellar Progenitor Metallcities}
\label{subsec:metal_theory}
The radioactive decay of $^{56}$Ni is the major energy source of the early time SN Ia light curve and the metallicity of the progenitor WD plays an important role in its production. \cite{timmesetal03} analytically showed that the yield of $^{56}$Ni drops with increased metallicity. Late-time light curves powered by the decay of $^{57}$Ni are also affected by metallicity. The ratio 57/56 increases with an increasing metallicity, because a metal-rich WD stellar progenitor has a lower electron fraction (i.e. a lower Y$_e$) and thus extra neutrons, as compared to a metal-poor WD stellar progenitor. This lower Y$_e$ shifts the nuclear statistical equilibrium from the dominant $^{56}$Ni yield to more neutron rich nuclides like $^{57}$Ni and $^{55}$Co.

The effect of progenitor metallicity can be seen in the sub-$M_{\rm{Ch}}$ models of \citet{shen2018sub} in Figure \ref{fig:subChar}, where increasing the metallicity for a fixed WD mass yields a higher value of the 57/56 ratio (darker shades have a higher 57/56 ratio) (See also Figures 28 and 29 in \citet{leung2020explosive}). This effect can also be seen in the near-$M_{\rm{Ch}}$ models of \citet{daveetal17} in Figure \ref{fig:densityLatetime} and in the models of \citet{leung2018explosive} in Figure \ref{fig:figureTest}a. (See also Figures 16 and 19 in \citet{leung2018explosive}).

In near-$M_{\rm{ch}}$ progenitors, the convective simmering phase \citep{chamulak2008reduction, piro2008neutronization} in the core of the WD sets a lower bound on the neutron excess, before the onset of a detonation. \citet{martinez2016neutronization} showed that the excess neutrons due to convective simmering is a very small fraction compared to the stellar progenitor metallicity for Z $\gtrsim$ 1/3 Z$_{\odot}$.

\subsection {Near-$M_{\rm{Ch}}$ WD Progenitors: Central Densities}
\label{subsec:central_rho_theory}
The central densities of WDs play an important role in the production of iron group elements in near-$M_{\rm{Ch}}$ progenitors. Progenitor WDs with higher central densities facilitate higher electron capture rates and thus higher neutronization. But higher central densities can indirectly reduce the level of neutronization - during the deflagration phase, higher densities allow for higher flame speeds. Due to high flame speed more fuel is burned and thus releases more energy, lead to a greater pre-expansion of the WD. This pre-expansion suppresses electron captures and degree of neutronization. 

The effect of the central density of a near-$M_{\rm{Ch}}$ WD on the neutronization of SN Ia yields are illustrated the gravitationally confined detonation (GCD) models presented by  \cite{daveetal17} in Figure \ref{fig:densityLatetime}. When the authors only varied the central density of the WDs with all other parameters fixed, the models with lower central density $\rho_c = 2.2 \times 10^9$ g cm$^{-3}$ produced a higher ratio of 57/56 as compared to a denser model with $\rho_c = 6 \times 10^9$ g cm$^{-3}$. This relation of central density and 57/56 ratio suggests a lower neutronization in a more centrally dense WD because of a higher degree of pre-expansion. Contrary to this trend, when \cite{daveetal17} reduced the C/O fraction to less than one, they found that the denser WD progenitors yielded more neutron rich isotopes because a lower carbon fraction reduces the flame speed leading to a lower energy release during deflagration. A lower energy release leads to a lower Atwood number and therefore less buoyancy and a lower growth exponential of the Rayleigh-Taylor instability. This causes the flame to spend more time at high density regions that facilitates more electron captures, and in turn produces a higher ratio of 57/56. 

Figure \ref{fig:figureTest}b shows DDT models with the same metallicity of Z=1.0Z$_{\odot}$ and equal carbon/oxygen ratio (C/O=1.0) from \citet{leung2018explosive} with central densities ranging from $1.0 \times 10^9$ g cm$^{-3}$ to $5.0 \times 10^9$ g cm$^{-3}$. In contrast to \citet{daveetal17} GCD models in Figure \ref{fig:densityLatetime}, higher central density produces a higher 57/56 ratio, even with C/O=1.0.


The central densities at the carbon ignition of the accreting WDs
depend on the accretion rate, the initial mass of the WD, and the
angular momentum of the WD \citep{nomoto82a, benvenuto2015final, nomoto2018single}.  Such high central densities as
$\rho_c = 5 - 6 \times 10^9$ g cm$^{-3}$ can be realized in the
following cases of accreting WDs.
\begin{enumerate}
    \item The accreting WD becomes cold if the accretion rate is so low that the radiative cooling exceeds the compressional heating.
    
    \item If the accretion rate is high, and if the WD is initially already cold and relatively massive, the central region remains cold until carbon ignition because there is not enough time for energy inflow from outer layers to heat up the central region of the WD.
    
    \item If the accreting WD gets enough angular momentum from the accretion disk materials to rotate fast and uniformly, the WD mass reaches $\sim 1.43~ M_{\odot}$ \citep{benvenuto2015final}, which is very close to $M_{\rm{Ch}}$, without igniting carbon due to the centrifugal force.  If such a rotating WD loses angular momentum, e.g., via magnetic winds, the WD contracts and ignites carbon at rather high central densities, Then carbon burning grows into thermonuclear runaway at $\rho_c = 5 - 6 \times$ 10$^9$ g cm$^{-3}$ as calculated by \cite{benvenuto2015final}.
\end{enumerate}

\subsection {Near-$M_{\rm{Ch}}$ WD Progenitors: Ignition}
\label{subsec:ignition_theory}
The ignition offset inside the near-$M_{\rm{Ch}}$ WD progenitor plays an important role in the level of neutronization in the SN Ia yields. The initialization of the ignition flame in the convective core of a near-$M_{\rm{Ch}}$ WD is stochastic in nature, with ignition offset going out to 100 km with an expectation value of 50 km, as shown with 3D simulations by \cite{zingale2011convective}.

With a single ignition bubble, \cite{fisher2015single} analytically showed that an ignition originating beyond 20 km from the center is buoyancy-driven and burns only a small fraction of the fuel as compared to a central ignition, which originates within 20 km of the center of the WD. This smaller energy release in the buoyancy-driven ignition leads to a smaller degree of pre-expansion, which results in the production of more neutronized isotopes during the detonation phase as compared to a centrally ignited deflagration flame, and thus leading to a higher ratio of 57/56.

The effect of ignition offset can be seen in the DDT models presented by \cite{daveetal17} in Figure \ref{fig:densityLatetime}, where the models with central density $\rho_c = 6 \times 10^9$ g cm$^{-3}$ and C/O = 30/70 are ignited centrally, i.e., with offset 0 km and others with a large offset of 200 km. The models with 0 km offset produce a lesser ratio of 57/56 as compared to models with offset 200km, which suggests a high degree of energy release during the deflagration phase in the centrally ignited models. This higher energy release leads to a higher degree of pre-expansion and less neutronization during the detonation phase as compared to the high offset models.

\section{Simulation Models and Observations}
\label{sec:lateTimeSimandObs}

The following section describes the various near-$M_{\rm{Ch}}$ and sub-$M_{\rm{Ch}}$ simulation models. The models presented vary greatly -- both in terms of the explosion channels simulated as well as in the numerical methods employed. The WDs also vary in the stellar progenitor metallicity and their masses, or equivalently their central densities. The detonation scenarios, primary WD central densities, and stellar progenitor metallicities are presented in Table \ref{tab:simruns}. 

We also present the estimates of $^{56}$Ni and $^{57}$Ni/$^{56}$Ni for the five SNe Ia, listed in the Table \ref{tab:obsTab} and a brief summary of how they were estimated from the late-time light curves.

\subsection{Simulations Models}
\label{sec:lateTimeSim}
Historically, near-$M_{\rm Ch}$ progenitors have been the favored model to explain SNe Ia. But recent observational evidence - absence of a main-sequence or red giant star in pre-explosion \citep{li2011exclusion,kelly2014constraints,graur2014progenitor,graur2019progenitor,do2021blast} and post-explosion data \citep{maoz2008search, edwards2012progenitor, hernandez2012no, kerzendorf2012hunting, schaefer2012absence, kerzendorf2014reconnaissance}, among others have led to a fall in favour of this model in the astrophysics community. In recent years, the sub-$M_{Ch}$ progenitor model has been gaining in popularity because the explosion of sub-$M_{Ch}$ WD can produce the light curves and spectra of normal SNe Ia \citep{sim2010detonations,maoz2014observational}. Moreover, the recent discovery of three hypervelocity WDs in the second release of the Gaia dataset \citep{shen2018three} supports the detonation of a sub-$M_{\rm Ch}$ WD. Moreover, the lack of hydrogen in the nebular phases pointed against the near-$M_{\rm Ch}$ progenitor \citet{leonard2007constraining}, but recent observations have found evidence for H-alpha lines in the late nebular phases for sub-luminous fast-declining SN Ia \citep{kollmeier2019h,vallely2019asassn,prieto2020variable,elias2021nebular}. Despite the fall in favour of the near-$M_{\rm Ch}$ channel, a few systems like SNR 3C 397 \citep{yamaguchi2015chandrasekhar,ohshiro2021discovery,zhou2021chemical} suggest strong observational evidence of a near-$M_{\rm Ch}$ progenitor.



We consider two SD explosion models: 
\begin{itemize}
    \item The deflagration to detonation transition (DDT) presented by \citet{daveetal17}, \citet{leung2018explosive}, \citet{seitenzahletal13b} and \citet{ohlmann2014white}. In the DDT model, the WD accretes from a main-sequence or a red giant star and increases its mass near the Chandrasekhar limit \citep{leung2018explosive}. The growth in mass raises the temperature and pressure near the core, which ignites a deflagration flame near the center of the WD, which in the process of rising to the surface transitions to a supersonic detonation and leads to a SN Ia. 
    \item The gravitational confined detonation (GCD) presented by \citet{daveetal17} and \citet{seitenzahletal16}.  In the GCD model, the deflagration doesn't transition to detonation, but rather rises to the surface and wraps around the WD, leading to the detonation of carbon after the merger of the deflagration fronts coming from the opposite sides.
\end{itemize}

We also consider:
\begin{itemize}
    \item Pure detonations presented in \citet{sim2010detonations} and \citet{shen2018sub} - in which a detonation is manually ignited in a sub-$M_{\rm Ch}$ WD.
    \item Violent mergers presented by \citet{Pakmor_2010}, \citet{pakmoretal12}, \citet{kromer2013sn} and \citet{kromeretal16} - where two sub-$M_{\rm Ch}$ WDs lose energy by gravitational radiation and would gradually come close enough for the secondary to be tidally disrupted and merge with the primary. This merger increases the density and temperature of the primary, leading to a SN Ia.
    \item The head-on collision of two sub-$M_{\rm Ch}$ WDs presented by \citet{papish2016supernovae} - where the eccentricity of a binary WD in a hierarchical triple system can be driven close to unity via Kozai-Lidov mechanism\citep{lidov1962evolution,kozai1962secular}, leading to a head-on collision. This collision will trigger shocks causing a thermonuclear explosion resulting in SN Ia \citep{katz2012rate,kushnir2013head}.
    \item The double detonation model presented by \citet{sim20122d}, \citet{Zen+18},  \citet{ZenatiPerets20} and \citet{leung2020explosive} - where the WDs have helium shells on their surfaces, and the detonation of the helium layer wraps around the primary and sends shock waves inwards, which leads to the detonation of carbon. We have omitted the models from \citet{leung2020explosive} with metallicities greater than 3 Z$_{\odot}$.
    \item We have omitted the double detonation models from \cite{sim20122d} whose WD CO progenitors with masses less than 0.58$M_{\odot}$ which yielded very low $\rm^{56}Ni$ ($\sim 0.1 M_{\odot}$), and therefore such models would produce very dim transients. We note that these double detonation models had very high mass ratios for 57/56 because of silicon burning \citep{woosley1973explosive}. As an example, the model with the maximum 57/56 = 0.17, yielded M($\rm^{56}Ni$)= 0.00282 M$_{\odot}$.
\end{itemize}


To model these different channels, the simulations usually follow a general pipeline - creating the initial conditions of a single WD using the WD density, temperature, composition profile either by evolving a main-sequence star to a WD in MESA \citep{paxton2010modules,paxton2013modules, paxton2015modules} or by using an appropriate equation of state of a hydrostatic WD. These profiles are then mapped to an Eulerian grid and evolved for that particular detonation model with tracer particles. These tracer particles are then post-processed through a large nuclear reaction network to get the final nucleosynthetic yields. Channels with binary WDs start by initializing and evolving a smooth particle hydrodynamics  simulation (SPH) of the system. The first snapshot that meets a certain condition (for example a threshold temperature is reached) is mapped onto the grid to follow the detonation fronts and tracer particles.

The Eulerian grid codes used in the simulation models are FLASH \citep{Fryxell_2000, dubeyetal09, dubeyetal13}, LEAFS \citep {reinecke1998new, reinecke2002refined}, PROMETHEUS \citep{fryxell1989mpa}, and the code presented by \cite{leung2015new}. The smooth particle hydrodynamics code in the merger models is GADGET-2 \citep{springel2005cosmological}. The nuclear reaction network codes used for post-processing are TORCH \citep{timmes99}, MESA's nuclear reaction module \citep{paxton2010modules,paxton2013modules, paxton2015modules} and the code presented in \citet{travaglio2004nucleosynthesis}.


\begin{table*}
	\centering
	  \begin{threeparttable}[t]
  \centering
        	\caption{Table of Simulation Runs.}
       \begin{tabular}{l   c  c  c}
    \hline
     Model-Dataset & Explosion Mechanism & $\rho_c$ (g cm$^{-3}$) & $Z/Z_\odot$ \\
    \hline
Dave et al, 2017 & near-$M_{\rm Ch}$-GCD & $2.2 \times 10^9 - 6 \times 10^9$  &  0-3 \\ 
Dave et al, 2017 & near-$M_{\rm Ch}$-DDT & $6 \times 10^9$  &  0-3 \\
Leung+Nomoto, 2018 & near-$M_{\rm Ch}$ DDT & $1 \times 10^9 - 5 \times 10^9$ & 0-5 \\
HESMA$^1$ & near-$M_{\rm Ch}$ GCD & $2.9 \times 10^9$ & 1 \\ 
HESMA$^1$ & near-$M_{\rm Ch}$ DDT & $1.0 \times 10^9 - 5.5 \times 10^9$ & 0.01-1 \\
HESMA$^1$ & sub-$M_{\rm Ch}$ Pure DET & $1.0 \times 10^7 - 2.0 \times 10^8$ & 0-3 \\
HESMA$^1$ & sub-$M_{\rm Ch}$ Merger & $2 \times 10^6 - 1.4 \times 10^7$ & 0.01 - 1 \\
HESMA$^1$ & sub-$M_{\rm Ch}$ Double Det & $2 \times 10^6 - 1.4 \times 10^7$ & 0.01 - 1 \\
Shen et al, 2018 & sub-$M_{\rm Ch}$ Pure DET & $\textnormal{N/A}$ & 0-2 \\
Papish+Perets, 2015 & sub-$M_{\rm Ch}$ Collision & $3.4 \times 10^6 - 1.0 \times 10^7$ &  0\\
Zenati+Perets, 2019 & sub-$M_{\rm Ch}$ Hybrid Double DET & $3.5 \times 10^6 -  1.2 \times 10^7$ &  1\\
Leung+Nomoto, 2020 & sub-$M_{\rm Ch}$ Double DET & $ 1.7 \times 10^7 - 1.5 \times 10^8$ &  0-5\\
     \hline
  \end{tabular}
     \begin{tablenotes}
     \item[1] HESMA - Heidelberg Supernova Model Archive - \citet {seitenzahletal13b} ,\citet{ ohlmann2014white} , \citet{seitenzahletal16} , \citet{sim2010detonations} , \citet{marquardt2015type} , \citet{Pakmor_2010} , \citet{pakmoretal12} , \citet{kromer2013sn} , \citet{kromeretal16} \cite{sim20122d}
   \end{tablenotes}
    \end{threeparttable}
	\label{tab:simruns}
\end{table*}

\subsection{Observations}
\label{sec:obs}

Determining the $^{57}$Ni/$^{56}$Ni ratios from the late-time light curves is quite challenging because multiple processes can produce late-time light curves tails. Processes like freeze-out\citep{fransson2015reconciling, kerzendorf2017extremely}, partial positron trapping\citep{dimitriadis2017late,kushnir2020constraints}, delayed decay of $^{56}$Ni or luminosity from a surviving companion\citep{shen2017wait} can mimic $^{57}$Co.

Light echoes have been observed to slow down the light curves of several SNe Ia starting at ~500 days (e.g., \citet{schmidt1994sn, sparks1999evolution, cappellaro2001detection, quinn2006light, wang2008detection, drozdov2015detection}), and so could mimic the late-time light curves of the SNe discussed here. This source of systematic uncertainty can be removed by studying the color evolution of the SNe. Because light echoes are created by the dispersion of the original SN light by dust sheets in space, the light that reaches our apertures will be bluer than the original SN light (e.g., \citet{patat2005reflections,rest2012light}). Furthermore, the light echo will be dominated by the light of the SN at peak, when it was most luminous. Hence, the colors of a SN contaminated by a light echo will be bluer than the colors of the SN at maximum light. The colors of SNe 2012cg, ASASSN-14lp, 2014J, and 2015F did not obey this rule; they were found to be bluer than peak in B-V but redder than peak in V-R \citep{graur2018observations,graur2018late_ASASSN,graur2018late,yang2018late}. The color evolution of these SNe, as well as SN 2013aa \citep{jacobson2018constraining} was also shown to be very similar to that of the well-studied SN 2011fe, which exploded in a very clean environment and has exhibited no light echoes \citep{graham2015constraining,shappeeetal16}. We note that SN 2014J has known, resolved light echoes \citep{crotts2015light,yang2017interstellar}, but \citet{yang2018late} and \citet{graur2018late} both note that the PSF of the SN within the aperture used for photometry is not contaminated by an unresolved echo.

Values of $^{56}$Ni and $^{57}$Ni/$^{56}$Ni for the five SNe Ia, listed in Table \ref{tab:obsTab}, are presented in the following subsection. \citet{childressetal15} estimated the mass of $^{56}$Ni for thirty one SNe Ia, including the five used in this work by looking at the [Co III] nebular spectral line of multiple SNe Ia, and showing that the line flux is proportional to the mass of $^{56}$Co at that time. \citet{childressetal15} note that [Co III] line flux measurement could be affected by contamination from neighboring nebular emission, undetected light echos, or residual host galaxy light. The decay of $^{56}$Co emits positrons and gamma rays. Under the assumption that positrons deposit their energy equally to all species in the ejecta and the ejecta being optically thin to gamma rays, \citet{childressetal15} show that the [Co III] line flux is proportional to the square of the mass of cobalt at that time. The estimated M($^{56}$Ni) for SN 2011fe, SN 2012cg, SN 2014J, SN 2015F and SN 2013aa are 0.500$^{+0.026}_{-0.026}$, 0.477$^{+0.048}_{-0.048}$, 0.838$^{+0.176}_{-0.176}$, 0.4$^{+0.05}_{-0.05}$ and 0.732$^{+0.151}_{-0.151}$ M$_{\odot}$ respectively.

The determination of $^{57}$Ni/$^{56}$Ni were made as follows: firstly the pseudo-bolometric light curves of the events in the optical range were calculated, using the observed fluxes from the \textit{HST}. These pseudo-bolometric fluxes were then converted to luminosities by taking the distances to the SNe Ia or the host galaxy into account. Finally, the Bateman equation \citep {seitenzahl2014light} is fit to the observed luminosities, to get the estimates for the yields and ratios of the nuclide. We would like to make a note that these observations are prone to systematic uncertainties, like missing multi-band observations for some SNe events, the uncertainties of the distance to the SNe would introduce errors to the calculated luminosities.

The pseudo-bolometric light curves of SNe 2012cg, 2013aa, ASASSN-14lp, 2014J, and 2015F cover the optical wavelength range of 3500 to 10000 A. \citet{shappeeetal16} had near-infrared (NIR) as well as optical observations of SN 2011fe, which they used to construct a pseudo-bolometric light curve spanning 4000 to 17000 A. 
\citet{graur2020year} showed that even though normal SNe Ia have varying light curve widths in the H band, their light curves converge after 500 days and follow the same decline rate as in the optical. The SNe in our sample cover a similar range of peak light curve widths (compare \citet{graur2018late_ASASSN} and \citet{graur2020year}). Hence, we assume that the SNe analysed here have the same NIR light curve shape at ~500-1000 days as in the optical. After 1000 days, only SN 2011fe has NIR observations \citep{shappeeetal16}. These mirror the slow-down of the optical light curves at the same phases. Since SN 2011fe is an archetypical normal SNe Ia and lies at the center of our range of peak light curve widths, we assume that the rest of the SNe in our sample had similar NIR light curves at >1000 days. This last point should be tested by NIR observations of future, nearby SNe Ia. Thus, while some care needs to be taken when producing pseudo-bolometric light curves in the phase range 150-500 days, it is safe to assume that during the late times studied here, $>$ 500 days past maximum light, the NIR and optical light curves behave in a similar manner.

\citet {shappeeetal16} used the light curve of SN 2011fe for 1840 days past maximum light and estimated log($^{57}$Ni/$^{56}$Ni) = -1.59$^{+0.06}_{-0.07}$ and also put an upper bound on the $^{55}$Fe/$^{57}$Co $<  0.22$, with 99\% confidence. Using these ratios, we place an upper bound of $^{55}$Fe/$^{56}$Co $< 0.006$ which we designate as 2011fe$^{\dagger}$.

\citet{tucker2021whisper} made the observations for SN 2011fe for $\approx$ 2400 days past maximum light and estimated log($^{57}$Ni/$^{56}$Ni) = -1.71$^{+0.12}_{-0.11}$ and made the first detection of $^{55}$Fe with log($^{55}$Fe/$^{57}$Ni)=-0.61$^{+0.13}_{-0.15}$, from which we estimated $^{55}$Fe/$^{56}$Ni = 0.00479$^{+0.00195}_{-0.00208}$, noted as 2011fe$^X$.

\citet {graur2016late} used the light curve of SN 2012cg for 1055 days past maximum light and found a mass ratio of M($^{57}$Co)/M($^{56}$Co) = 0.043$^{+0.012}_{-0.011}$, noted as 2012cg$^{\dagger}$.

\citet{li2019observations} followed the normal SN Ia 2014J to $\sim 900$ days after maximum light, and from its late-time light curve fit for $^{56}$Ni, $^{57}$Ni, and $^{55}$Fe to get the $^{57}$Ni/$^{56}$Ni mass ratio, which they estimated to be 0.035$^{+0.011}_{-0.011}$. They took the yields of the three isotopes as free parameters and accounted for lepton escape to estimate M($^{56}$Ni) = 0.68$^{+0.12}_{-0.12}$ $M_{\odot}$. Their $^{56}$Ni yields are consistent with the estimates by \citet{srivastav2016optical}, \citet{telesco2014mid}, \citet{churazov2014cobalt}, but are in contrast with the yield calculated by \citet{childressetal15}, which is M($^{56}$Ni) = 0.838$^{+0.176}_{-0.176}$ M$_{\odot}$. They note that their $^{57}$Ni/$^{56}$Ni estimates are half that of \citet{yang2018mapping}, as they did not take lepton escape into account.

\citet{graur2018observations} fit the pseudo-bolometric light curves of SN 2011fe, SN 2012cg, SN 2014J and SN 2015F but only considered the contributions from $^{56}$Ni and $^{57}$Ni to estimate the $^{57}$Ni/$^{56}$Ni ratios as 0.043$^{+0.004}_{-0.004}$ (2011fe$^*$), 0.072$^{+0.002}_{-0.002}$ (2012cg$^*$), 0.129$^{+0.021}_{-0.016}$ (2014J$^*$) and 0.004${^{+0.003}_{-0.002}}$ (2015F), respectively.

\citet{jacobson2018constraining} combined archival data at $< 400$ days with one epoch of \textit{HST} observation at 1500 days after explosion of the over-luminous SN 2013aa. They estimated the mass of $^{56}$Ni = 0.732$^{+0.151}_{-0.151}$ M$_{\odot}$ by fitting a straight line to the M($^{56}$Ni) values over SiFTO stretch values by \citet{childressetal15}. To estimate the mass ratio M($^{57}$Co)/M($^{56}$Co), they fixed the estimated M($^{56}$Ni) and found a mass ratio M($^{57}$Co)/M($^{56}$Co) = 0.02$^{+0.01}_{-0.02}$ (2013aa$^*$). The authors also allowed for variation in $^{56}$Ni for which they estimated the M($^{56}$Ni) = 0.631$^{+0.015}_{-0.015}$ M$_{\odot}$, and from their estimates of M($^{57}$Co) and M($^{56}$Co) we compute the M($^{57}$Co)/M($^{56}$Co) = 0.0095$^{+0.00016}_{-0.00095}$ (2013aa$^{\dagger}$).

\begin{table*}
\begin{center}
	\centering
	  \begin{threeparttable}[ht]
  \centering
  \caption{\label{tab:obsTab}Event Types}
       \begin{tabular}{lccccccc}
    \hline
    Event & & & M($^{56}$Ni) & & $^{57}$Ni/$^{56}$Ni & & Wavelength Range (\AA)\\
    \hline
2011fe & & & 0.500$^{+0.026}_{-0.026}$ & 0.043$^{+0.004*}_{-0.004}$ & 0.026$^{+0.0037\dagger}_{-0.0044}$ & 0.0195$^{+0.0054X}_{-0.0049}$ & 4000-17000 \\ 
2012cg & & & 0.477$^{+0.048}_{-0.048}$ & 0.072$^{+0.002*}_{-0.002}$ & 0.043$^{+0.011\dagger}_{-0.012}$ & & 3500-10000\\
2014J & & & 0.838$^{+0.176}_{-0.176}$ & 0.129$^{+0.021*}_{-0.016}$ & 0.035$^{+0.011\dagger}_{-0.011}$ & & 3500-10000\\ 
2015F & & & 0.400$^{+0.05}_{-0.05}$ & 0.004$^{+0.003}_{-0.002}$ & & & 3500-10000\\
2013aa & & & 0.732$^{+0.151}_{-0.151}$ & 0.02$^{+0.01*}_{-0.02}$ & 0.0095$^{+0.0016\dagger}_{-0.0095}$ & & 3500-10000\\
     \hline
  \end{tabular}
     \begin{tablenotes}
     \item All of the SNe are normal, but that 2015F is subluminous and 2013aa is overluminous. Events with multiple estimates of $^{57}$Ni/$^{56}$Ni are marked with superscript dagger($\dagger$) or asterisk($*$) or cross($X$). \\
        2014J$^{\dagger}$ - \citet{li2019observations} 
		2011fe$^{\dagger}$ - \citet{shappeeetal16}, 
		2012cg$^{\dagger}$ - \citet{graur2016late},
		\\2014J$^{*}$, 2011fe$^{*}$, 2012cg$^{*}$, 2015F - \citet{graur2018observations}
		\\2013aa$^{*}$-Fit 3 , 2013aa$^{\dagger}$-Fit 2 \cite{jacobson2018constraining}
		\\2011fe$^{X}$ - \cite{tucker2021whisper}
   \end{tablenotes}
    \end{threeparttable}
	\end{center}
\end{table*}

\section{Results}
\label{sec:LateTimeresults}

\subsection {WD Progenitors: sub-$M_{\rm Ch}$ and near-$M_{\rm Ch}$}

\begin{figure*}
    \includegraphics[width=1.05\textwidth]{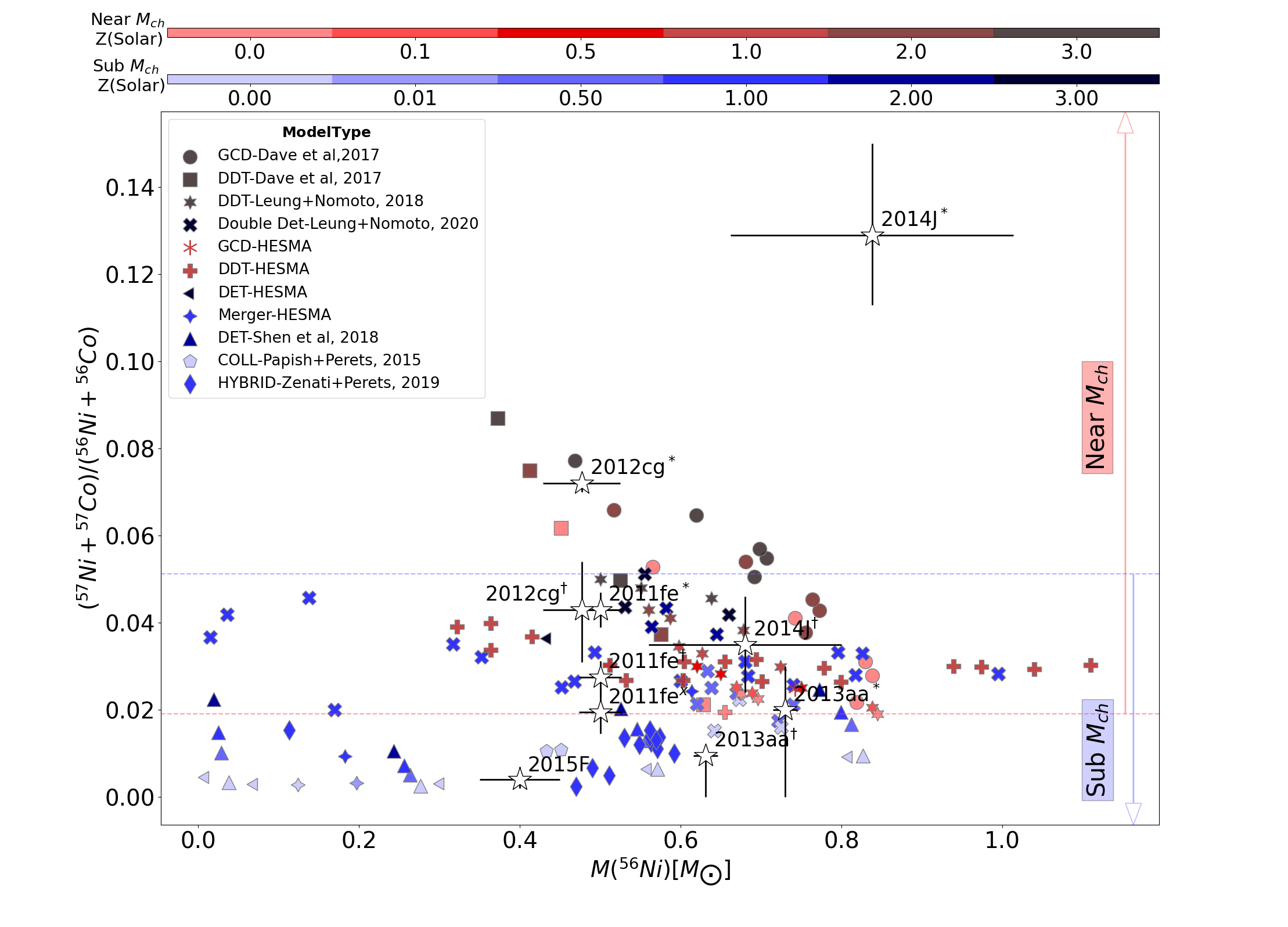}
	\caption{Plot showing the ratio ($\rm{^{57}Ni}$+$\rm{^{57}Co}$)/($\rm{^{56}Ni}$+$\rm{^{56}Co}$) against $M(\rm{^{56}Ni})$. The horizontal blue line marks the upper bound on 57/56 ratio among all the sub-$M_{\rm{Ch}}$ models, and the red horizontal line shows the lower bound on all the near-$M_{\rm{Ch}}$ models considered in this work. The near-$M_{\rm Ch}$ models(red) have on average much higher 57/56 ratios due to their higher central densities as compared to the sub-$M_{\rm Ch}$(blue) progenitors.  An overlap region exists between the two channels, where sub-$M_{\rm Ch}$ WD models with supersolar progenitor metalicities can have neutronizations comparable to those of near-$M_{\rm Ch}$ models. Only sub-$M_{\rm{Ch}}$ models are consistent with event SN 2015F and SN 2011fe$^{X}$.
	Only near-$M_{\rm{Ch}}$ models are consistent with SN 2012cg$^*$. The events SN 2012cg$^{\dagger}$, 2012cg$^{*}$, 2011fe$^{\dagger}$, 2014J$^{\dagger}$, 2013aa$^{*}$, 2013aa$^{\dagger}$ match with both the near-$M_{\rm{Ch}}$ and sub-$M_{\rm{Ch}}$ models. 
	\label{fig:57by56plot}} 
\end{figure*}

To place constraints on the simulation models, we plot the abundance ratio of the 57 isobar chain to the 56 isobar chain ($\rm{^{57}Ni}$+$\rm{^{57}Co}$)/($\rm{^{56}Ni}$+$\rm{^{56}Co}$) against $M(\rm{^{56}Ni})$ in Figure \ref{fig:57by56plot}. The plot shows the near-$M_{\rm Ch}$ models in red and the sub-$M_{\rm Ch}$ models in blue, with different shades representing their metallicities. As we go up the ($\rm{^{57}Ni}$+$\rm{^{57}Co}$)/($\rm{^{56}Ni}$+$\rm{^{56}Co}$) axis the degree of neutronized nuclides increases while going towards the increasing $\rm^{56}Ni$ yields on the x-axis we would expect brighter events. Further, we expect to see more neutronized isotopes in a much denser environment because of higher electron captures, which are the signature of dense near-$M_{\rm Ch}$ progenitors. From Figure \ref{fig:57by56plot}, we see that the near-$M_{\rm Ch}$ models have on average much higher 57/56 ratios due to their higher central densities as compared to the sub-$M_{\rm Ch}$ progenitors.  An overlap region exists between the two channels, where sub-$M_{\rm Ch}$ WD models with supersolar progenitor metalicities can have neutronizations comparable to those of near-$M_{\rm Ch}$ models. This shows how the late-time light curves can be used as a new probe to constrain the parameters of the progenitors of SNe Ia.


\begin{figure*}
	\includegraphics[width=1.05\textwidth]{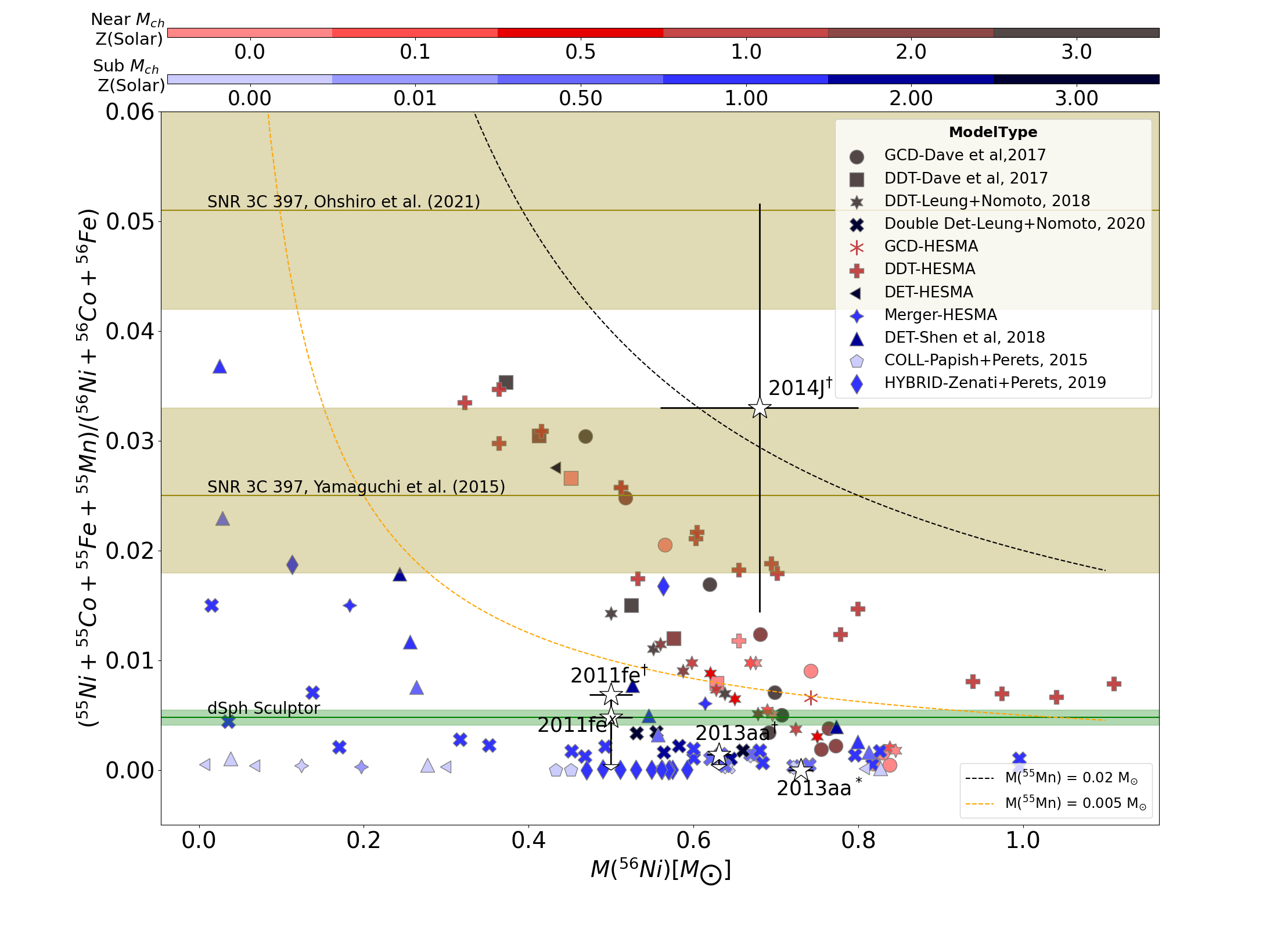}
	\caption{($\rm{^{55}Ni}$+$\rm{^{55}Co}$+$\rm{^{55}Fe}$)/($\rm{^{56}Ni}$+$\rm{^{56}Co}$) versus $M(\rm{^{56}Ni}$) for near-$M_{\rm{Ch}}$(red) and sub-$M_{\rm{Ch}}$(blue) models, with the event 2011fe$^X$, 2014J$^{\dagger}$ and upper bounds on the events 2011fe$^{\dagger}$, 2013aa$^*$ and 2013aa$^{\dagger}$. The black and yellow dashed lines show the region where the yields of $\rm^{55}Mn$ are in the range of 0.002 M$_{\odot}$ to 0.005 M$_{\odot}$; most of the near-$M_{\rm{Ch}}$ models fall in this range. The mass ratio of Mn/Fe of the SN remnant SNR 3C 397 is shown as the orange bar, which supports a near-$\rm M_{Ch}$ progenitor for this SNR. To validate the potential of using the late-time light curve channel, we also plot the values for Mn/Fe from \citet{de2020manganese} for the dSph Sculptor galaxy, shown as the green bar. \label{fig:55by56plot}}
\end{figure*}

Further, one can also note the relation of metallicities in the \cite {daveetal17} DDTs and GCDs, in which more metal-rich WDs have a higher 57/56 ratio and lower yields of $^{56}$Ni. Similar trends can be seen in \cite {leung2018explosive} DDTs and \cite {shen2018sub} pure detonations of sub-$M_{\rm Ch}$ WDs.

We qualitatively compare the five near by observations with the models by plotting them in Figure \ref{fig:57by56plot}. We note that the events 2011fe, 2012cg and 2014J each have multiple estimates, the first one calculated by fitting the Bateman equation using the nuclides $\rm^{56}Ni$, $\rm^{57}Ni$ and $\rm^{55}Co$, which are all marked with daggers, while the other estimates were computed using only the nuclides $\rm^{56}Ni$ and $\rm^{57}Ni$ are marked with asterisk. SN 2011fe has a third estimate from very late-time light curve marked as cross. We can see that only SN 2015F and SN 2013aa$^{\dagger}$ are in the ``only sub-$\rm M_{Ch}$ region'' i.e. in the region where 57/56 $\lessapprox$ 0.02, while 2012cg$^{*}$ and 2014J$^*$ lie in the ``near-$\rm M_{Ch}$'' region, i.e. where 57/56 $\gtrapprox$ 0.05. SN 2012cg$^{\dagger}$, 2011fe$^*$, 2011fe$^{\dagger}$, 2011fe$^{X}$, 2014J$^{\dagger}$ and 2013aa$^*$ lie in the overlap region.

To quantitatively compare the simulation models with the observations we perform a $\chi^2$ test on each of the observations to find the set of models which are consistent with them using:

\begin{equation}
\label{chisq}
\begin{split}
\chi^2 = \{ \frac{(X^{57/56}_{i,obs} - X^{57/56}_{j, model})^{2}}{\sigma_{i, 57/56}^{2}} + \frac{(X^{56}_{i,obs} - X^{56}_{j,model})^{2}}{\sigma_{i,56}^{2}} \\
  + \frac{(X^{55/56}_{i,obs} - X^{55/56}_{j, model})^{2}}{\sigma_{i, 55/56}^{2}} \}
\end{split}
\end{equation}

and summarize the results in Table \ref{tab:long}. We use the upper-tail one-sided $\chi^2$ test with the null hypothesis defined as ``the event being consistent with a simulation model'', and select a significance value of 0.001. An event being consistent with a model means that the model can explain the observed late-time light curve of the SN event.

From the analysis we see SN 2012cg$^*$ is consistent with 4 near-$M_{\rm Ch}$ DDT and GCD models of \citet{daveetal17}. SN 2011fe$^X$ is only consistent with sub-$M_{\rm Ch}$ models of \citet{shen2018sub}, \citet{leung2020explosive}, \citet{ZenatiPerets20} and \citet{papish2016supernovae}. SN 2015F can only we explained by sub-$M_{\rm Ch}$ collision models of \citet{papish2016supernovae} and sub-$M_{\rm Ch}$ models of \citet{ZenatiPerets20}.

We find that the observational event SN 2011fe$^*$ is consistent with 11 near-$M_{\rm Ch}$ DDT and GCD models of \citet{seitenzahletal13b}, \citet{daveetal17} and \citet{leung2018explosive} and 6 sub-$M_{\rm Ch}$ double detonation models of \citet{leung2020explosive}. SN 2011fe$^{\dagger}$ is consistent with 20 sub-$M_{\rm Ch}$ models from \citet{leung2020explosive}, \citet{shen2018sub}, \citet{ZenatiPerets20}, \citet{papish2016supernovae} and \citet{sim2010detonations} and also consistent with 4 near-$M_{\rm Ch}$ models of \citet{daveetal17} and \citet{seitenzahletal13b}.

For SN 2012cg$^{\dagger}$, 7 near-$M_{\rm Ch}$ DDT and GCD models of \citet{seitenzahletal13b}, \citet{daveetal17} and \citet{leung2018explosive} and 7 sub-$M_{\rm Ch}$ double detonation and pure detonation models of \citet{leung2020explosive} and \citet{sim2010detonations} can explain the late-time light curve.

We find SN 2014J$^{\dagger}$ to be consistent with 25 near-$M_{\rm Ch}$ models of HESMA \citep{seitenzahletal13b, ohlmann2014white, seitenzahletal16}, \citet{daveetal17} and \citet{leung2018explosive}, and also to be consistent with 17 sub-$M_{\rm Ch}$ double detonation models from \citep{leung2018explosive} and a violent merger model of \citet{pakmoretal12}. We note that SN 2014J$^{*}$ is not consistent with any simulation model.

For SN 2013aa$^{*}$, 16 near-$M_{\rm Ch}$ DDT and GCD models of HESMA \citep{seitenzahletal13b, ohlmann2014white, seitenzahletal16}, \citet{daveetal17} and \citet{leung2018explosive} and 18 sub-$M_{\rm Ch}$ double detonation and pure detonation models of \citet{leung2020explosive}, \cite{shen2018sub} and \citet{sim2010detonations} are consistent with the measurement. For SN 2013aa$^{\dagger}$, 23 sub-$M_{\rm Ch}$ progenitor models of \citet{ZenatiPerets20}, \citet{leung2020explosive} and \citet{shen2018sub} and 9 near-$M_{\rm Ch}$ DDT models of HESMA(\citet{seitenzahletal13b}, \citet{ohlmann2014white}), \citet{daveetal17} and \cite{leung2018explosive} are consistent with the measurement.

We can also explore constraints on the events using the very long-lived $A=55$ isobar chains, which have half-lives of $\sim$ 3 yrs. However, due to the difficulty of observing SNe Ia at very late times, estimates of $M(\rm{^{55}Co}$)/$M(\rm{^{56}Co}$) have been made for only a few events. \cite {shappeeetal16} put an upper limit on the mass ratio of 55/57 chain for SN 2011fe$^{\dagger}$ ($6.6 \times 10^{-3}$) with 99\% confidence and \cite{jacobson2018constraining} estimated M($^{55}$Fe) using which we put an upper bound for 55/56 for SN 2013aa$^{\dagger}$ $(1.42 \times 10^{-3}$) and SN 2013aa$^{*}$ ($1.045 \times 10^{-5}$). \citet{tucker2021whisper} estimated log($^{55}$Fe/$^{57}$Ni)=-0.61$^{+0.13}_{-0.15}$. To use this channel, we plot ($\rm{^{55}Ni}$+$\rm{^{55}Co}$+$\rm{^{55}Fe}$)/($\rm{^{56}Ni}$+$\rm{^{56}Co}$) versus $M(\rm{^{56}Ni}$) for the simulation models in Figure \ref{fig:55by56plot}. The black and yellow dashed lines show the region where the yields of $\rm^{55}Mn$ are in the range of 0.002 M$_{\odot}$ to 0.005 M$_{\odot}$, and we see that most of the near-$M_{Ch}$ models fall in this range. We also show the upper bounds for SN 2011fe$^{\dagger}$, SN 2013aa$^{\dagger}$ and SN 2013aa$^*$. We don't use the 55/56 bound for SN 2013aa$^*$ in our analysis because the value and error bar on 55/56 are very close to zero.

Further, we include results of the mass ratio of Mn/Fe of the remnant SNR 3C 397, which indicates a near-$\rm M_{Ch}$ progenitor. \cite{yamaguchi2015chandrasekhar} and \cite{ohshiro2021discovery} determined the ratio of Mn/Fe for 3C 397 to be in the range 0.018–0.033 and 0.042-0.060, respectively, which is shown as a shaded yellow bar in Figure \ref{fig:55by56plot}. \cite{zhou2021chemical} also determined the mass ratio of Mn/Fe of SNR Sgr A East to be in the range of 0.023 - 0.032, being similar to 3C 397, and concluded that the progenitor of Sgr A East is the near-$\rm M_{Ch}$ white dwarf. Qualitatively looking at the Figure \ref{fig:55by56plot}, we find that SNR 3C 397 likely had a near-$\rm M_{Ch}$ progenitor, which is consistent with the constraint put by \cite{yamaguchi2015chandrasekhar,ohshiro2021discovery,zhou2021chemical}. To further validate the potential of using the late-time light curve channel, we plot the the estimates for Mn/Fe from \citet{de2020manganese} for the dSph Sculptor galaxy, where they estimated the Mn/Fe contributions from SNe Ia and used that to put constraints on nature of the SNe Ia progenitor. The authors found that the major contribution to Mn/Fe in the dSph Sculptor galaxy was from sub-$\rm M_{Ch}$ progenitors. We plot their estimates in our Figure \ref{fig:55by56plot} as the green shaded region, and also find their estimates to be consistent with sub-$\rm M_{Ch}$ progenitors. We also note that \citet{de2020manganese} noted that some fraction (> 20 \%) of near-$\rm M_{Ch}$ progenitors are still needed to produce the observed yield. They further suggested that near-$\rm M_{Ch}$ progenitors may become the dominant channel of SNe Ia at later times in a chemical evolution of the dSph Sculptor galaxy.

\begin{table*}
	\centering
	\caption{Models Consistent  with Events.}
	\label{tab:long}
	  \begin{threeparttable}[ht]
  \centering
       \resizebox{\textwidth}{!}{\begin{tabular}{|l|c|c|c|c|}
    \hline
    Observation & Model & Number of Models & Z$_{\odot}$ & Parameters\\
    \hline
2011fe$^{*}$ & near-$M_{\rm Ch}$ DDT - Leung+Nomoto, 2018 & 4 & 2.0-3.0 & $\rho$: 3e9 - 5e9 g cm$^{-3}$, c3 flame - C/O:1  \\\cline{2-5}
 & near-$M_{\rm Ch}$ DDT - HESMA$^{\rm a}$-Seitenzahl et al.(2013b) & 4 & 1.0 & $\rho$: 1e9 - 2.9e9 g cm$^{-3}$ , N:100, 150, 200, 300 \\\cline{2-5}
 & near-$M_{\rm Ch}$ DDT,GCD - Dave et al, 2017 & 3 & 0.0-3.0 & $\rho$: 6.0e9 g cm$^{-3}$, bubble offset: 0km - 200km, bubble radius: 8km - 100km, C/O:30/70 \\\cline{2-5}
 & sub-$M_{\rm Ch}$ Double DET - Leung+Nomoto, 2020 & 6 & 1.0-3.0 & $\rho$: 4.33e7 - 6.17e7 g cm$^{-3}$, Helium Detonation: Spherical, Bubble, Ring. \\
\hline
2011fe$^{\dagger}$ & sub-$M_{\rm Ch}$ Double DET - Leung+Nomoto, 2020 & 8 & 1.0-3.0 & $\rho$: 2.26e7 - 6.17e7 g cm$^{-3}$, Helium Detonation: Spherical, Bubble, Ring. \\\cline{2-5}
 & sub-$M_{\rm Ch}$ Pure DET - Shen et al, 2018 & 3 & 0.5-2.0 & C/O = 50/50, $WD_{\rm{Mass}}$: 1.0 \\\cline{2-5}
 & sub-$M_{\rm Ch}$ Hybrid Double DET - Zenati+Perets, 2019 & 7 & 1.0 & M$_{\rm Hybrid}$: 0.53 M$_{\odot}$ -  0.68 M$_{\odot}$ , M$_{\rm CO}$ = 0.8 M$_{\odot}$ - 1.0 M$_{\odot}$ \\\cline{2-5}
 & sub-$M_{\rm Ch}$ Collision - Papish+Perets, 2015 & 1 & 0.0 & M$_{\rm WD1}$: 0.6 M$_{\odot}$ CO + 0.01 M$_{\odot}$ He - $\rho_1$: 3.4e6 g cm$^{-3}$ - M$_{\rm WD2}$: 0.6 M$_{\odot}$ CO + 0.01 M$_{\odot}$ He - $\rho_2$: 3.4e6 g/cm$^3$ \\\cline{2-5}
 & sub-$M_{\rm Ch}$ Pure DET - HESMA$^{\rm a}$-Sim et al. (2010) & 1 & 3.0 & $\rho$: 4.15e7 g cm$^{-3}$, M$_{\rm WD}$: 1.06 M$_{\odot}$ \\\cline{2-5}
 & near-$M_{\rm Ch}$ DDT - HESMA$^{\rm a}$-Seitenzahl et al.(2013b) & 3 & 1.0 & $\rho$: 1.0e9 - 2.0e9 g cm$^{-3}$ , N:100, 150, 300 \\\cline{2-5}
 & near-$M_{\rm Ch}$ DDT - Dave et al, 2017 & 1 & 1.5 & $\rho$: 6.0e9 g cm$^{-3}$, bubble offset:0 km - bubble radius:100km - C/O:30/70 \\
\hline
2011fe$^{X}$ & sub-$M_{\rm Ch}$ Pure DET - Shen et al, 2018 & 3 & 0.5-2.0 & C/O = 50/50, $WD_{\rm{Mass}}$: 1.0 \\\cline{2-5}
& sub-$M_{\rm Ch}$ Double DET - Leung+Nomoto, 2020 & 3 & 1.0 & $\rho$: 2.26e7 - 4.33e7 g cm$^{-3}$, Helium Detonation: Spherical, Bubble, Ring. \\\cline{2-5}
& sub-$M_{\rm Ch}$ Hybrid Double DET - Zenati+Perets, 2019 & 9 & 1.0 & M$_{\rm Hybrid}$: 0.41 M$_{\odot}$ -  0.73 M$_{\odot}$ , M$_{\rm CO}$ = 0.7 M$_{\odot}$ - 1.0 M$_{\odot}$ \\\cline{2-5}
& sub-$M_{\rm Ch}$ Collision - Papish+Perets, 2015 & 2 & 0.0 & M$_{\rm WD1}$: 0.6 M$_{\odot}$ CO + 0.0-0.01 M$_{\odot}$ He - $\rho_1$: 3.4e6 g cm$^{-3}$ - M$_{\rm WD2}$: 0.6 M$_{\odot}$ CO + 0.0-0.01 M$_{\odot}$ He - $\rho_2$: 3.4e6 g/cm$^3$ \\\cline{2-5}
\hline
2012cg$^{*}$ & near-$M_{\rm Ch}$ DDT,GCD - Dave et al, 2017 & 4 & 0.0-3.0 & $\rho$: 6.0e9 g cm$^{-3}$, bubble offset:200 km - bubble radius:8km - C/O:30/70. \\\cline{2-5}
 \hline
2012cg$^{\dagger}$ & near-$M_{\rm Ch}$ DDT - Leung+Nomoto, 2018 & 3 & 2.0-3.0 & $\rho$: 5.0e9 g cm$^{-3}$, c3 flame - C/O:1. \\\cline{2-5}
 & near-$M_{\rm Ch}$ DDT - HESMA$^{\rm a}$ - Seitenzahl et al.(2013b) & 3 & 1.0 & $\rho$: 2.9.0e9 g cm$^{-3}$, N: 150,200,300. \\\cline{2-5}
 & near-$M_{\rm Ch}$ DDT - Dave et al, 2017 & 1 & 3.0 & $\rho$: 6.0e9 g cm$^{-3}$, bubble offset:0km, bubble radius:100km, C/O:30/70. \\\cline{2-5}
 & sub-$M_{\rm Ch}$ Double DET - Leung+Nomoto, 2020 & 6 & 1.0-3.0 & $\rho$: 2.26e7 - 6.17e7 g cm$^{-3}$, Helium Detonation: Spherical, Bubble, Ring. \\\cline{2-5}
  & sub-$M_{\rm Ch}$ Pure DET - HESMA$^{\rm a}$-Sim et al. (2010) & 1 & 3.0 & $\rho$: 4.15e7 g cm$^{-3}$, M$_{\rm WD}$: 1.06 M$_{\odot}$ \\
\hline
2014J$^{*}$ & \textbf{No Models are Consistent with the Measurement} &  &  &  \\
\hline
2014J$^{\dagger}$ & near-$M_{\rm Ch}$ DDT,GCD - HESMA$^{\rm a}$  & 7 & 0.01-1.0 & $\rho$: 2.9e9 - 5.5e9 g cm$^{-3}$, DDT Models N:40,100 ,  GCD model Offset: 200 km. \\\cline{2-5}
 & near-$M_{\rm Ch}$ DDT - Leung+Nomoto, 2018 & 13 & 0.0-3.0 & $\rho$: 1.0e9 - 5.0e9 g cm$^{-3}$, c3 flame - C/O:1. \\\cline{2-5}
 & near-$M_{\rm Ch}$ DDT - Dave et al, 2017 & 5 & 0.0-3.0 & $\rho$: 2.2e9 - 6.0e9 g cm$^{-3}$, bubble offset:0km -100 km, bubble radius: 8km - 100km, C/O:30/70 , 50/50. \\\cline{2-5}
  & sub-$M_{\rm Ch}$ Double DET - Leung+Nomoto, 2020 & 16 & 0.0-3.0 & $\rho$: 3.21e7 - 6.17e7 g cm$^{-3}$, Helium Detonation: Spherical, Bubble, Ring. \\\cline{2-5}
  & sub-$M_{\rm Ch}$ Merger - Pakmor et al, 2012b & 1 & 1.0 & M$_{\rm WD1}$: 0.9 M$_{\odot}$, M$_{\rm WD2}$: 1.1 M$_{\odot}$ \\
\hline
2015F & sub-$M_{\rm Ch}$ Collision - Papish+Perets, 2015 & 2 & 0.0 & M$_{\rm WD1}$: 0.6 M$_{\odot}$ CO + 0.0-0.01 M$_{\odot}$ He - $\rho_1$: 3.4e6 g cm$^{-3}$ - M$_{\rm WD2}$: 0.6 M$_{\odot}$ CO + 0.0-0.01 M$_{\odot}$ He - $\rho_2$: 3.4e6 g/cm$^3$ \\\cline{2-5}
& sub-$M_{\rm Ch}$ Hybrid Double DET - Zenati+Perets, 2019 & 2 & 1.0 & M$_{\rm Hybrid}$: 0.41 M$_{\odot}$ -  0.48 M$_{\odot}$ , M$_{\rm CO}$ = 0.6 M$_{\odot}$ - 0.7 M$_{\odot}$ \\\cline{2-5}
\hline
2013aa$^{*}$ & near-$M_{\rm Ch}$ DDT - Leung+Nomoto, 2018 & 7 & 0.0-1.0 & $\rho$: 1.0e9 - 5.0e9 g cm$^{-3}$, c3 flame - C/O:1. \\\cline{2-5}
 & near-$M_{\rm Ch}$ DDT,GCD - HESMA$^{\rm a}$  & 7 & 0.01-1.0 & $\rho$: 2.9e9 - 5.5e9 g cm$^{-3}$, DDT Models N:20,40,100 ,  GCD model Offset: 200 km. \\\cline{2-5}
 & near-$M_{\rm Ch}$ GCD - Dave et al, 2017 & 2 & 0.0-1.5 & $\rho$: 6.0e9 g cm$^{-3}$, bubble offset: 100 km, bubble radius: 16km, C/O:50/50. \\\cline{2-5}
 & sub-$M_{\rm Ch}$ Double DET - Leung+Nomoto, 2020 & 14 & 0.0-1.0 & $\rho$: 4.33e7 - 9.19e7 g cm$^{-3}$, Helium Detonation: Spherical, Bubble, Ring. \\\cline{2-5}
 & sub-$M_{\rm Ch}$ Pure DET - Shen et al, 2018 & 3 & 0.5-2.0 & C/O = 50/50, M$_{WD}$: 1.1 M$_{\odot}$. \\\cline{2-5}
 & sub-$M_{\rm Ch}$ Pure DET - HESMA$^{\rm a}$ - Sim et al. (2010) & 1 & 0.0 & $\rho$: 7.9e7 g cm$^{-3}$, M$_{\rm WD}$: 1.15 M$_{\odot}$ \\
\hline
2013aa$^{\dagger}$ & sub-$M_{\rm Ch}$ DDT - Leung+Nomoto, 2020 & 11 & 0.0-1.0 & $\rho$: 3.21e7 - 6.17e7 g cm$^{-3}$, Spherical Det, Single He detonation, Ring Detonation. \\\cline{2-5}
 & sub-$M_{\rm Ch}$ Hybrid Double DET - Zenati+Perets, 2019 & 9 & 1.0 & M$_{\rm Hybrid}$: 0.53 M$_{\odot}$ -  0.74 M$_{\odot}$ , M$_{\rm CO}$ = 0.8 M$_{\odot}$ - 1.0 M$_{\odot}$ \\\cline{2-5}
 & sub-$M_{\rm Ch}$ Pure DET - Shen et al, 2018 & 3 & 0.0-1.0 & C/O = 50/50, M$_{WD}$: 1.0 M$_{\odot}$. \\\cline{2-5}
 & near-$M_{\rm Ch}$ DDT - HESMA$^{\rm a}$  & 2 & 0.01-1.0 & $\rho$: 2.9e9 g cm$^{-3}$, DDT Models N:100. \\\cline{2-5}
 & near-$M_{\rm Ch}$ DDT - Leung+Nomoto, 2018 & 6 & 0.0-0.5 & $\rho$: 3.0e9 - 5.0e9 g cm$^{-3}$, c3 flame - C/O:1. \\\cline{2-5}
 & near-$M_{\rm Ch}$ DDT - Dave et al, 2017 & 1 & 0.0 & $\rho$: 6.0e9 g cm$^{-3}$, bubble offset: 0 km, bubble radius: 100km, C/O:30/70. \\\cline{2-5}
     \hline
  \end{tabular}}
     \begin{tablenotes}
    \item HESMA - Heidelberg Supernova Model Archive \\
    2014J$^{\dagger}$ - \citet{li2019observations}, 2011fe$^{\dagger}$ - \citet{shappeeetal16}, 2012cg$^{\dagger}$ - \citet{graur2016late},\\
    2014J$^{*}$, 2011fe$^{*}$, 2012cg$^{*}$, 2015F - \citet{graur2018observations}\\
    2013aa$^{*}$-Fit 3 , 2013aa$^{\dagger}$-Fit 2 \cite{jacobson2018constraining}\\
    2011fe$^X$ - \citet{tucker2021whisper}
   \end{tablenotes}
    \end{threeparttable}
	\label{tab:simruns}
\end{table*}

\subsection {Stellar Progenitor Metallcities}

As discussed in section \ref{subsec:metal_theory}, the yield of $^{56}$Ni drops with increased metallicity. This effect can be seen in the sub-$M_{\rm{Ch}}$ models presented by \cite{shen2018sub}, and also in the near-$M_{\rm{Ch}}$ models presented by \cite{daveetal17} and \cite{leung2018explosive}.

In section \ref{subsec:metal_theory}, we also discussed how the ratio 57/56 increases with an increasing metallicity. This effect can be seen in the models presented by \cite{shen2018sub}, \cite{daveetal17} and \cite{leung2018explosive} in Figure \ref{fig:subChar} -- \ref{fig:57by56plot}.

In our analysis, we find that the models consistent with the events SN 2011fe$^*$, 2011fe$^{\dagger}$, 2012cg$^*$ and 2014J$^{\dagger}$ have metallcities ranging from 0.0 Z$_{\odot}$ - 3.0 Z$_{\odot}$, while SN 2012cg$^{\dagger}$ has metallicitiy in the range 1.0 Z$_{\odot}$ - 3.0 Z$_{\odot}$.
The consistent models for SN 2011fe$^{X}$ and 2013aa$^*$ have metallcities in the range 0.0 Z$_{\odot}$ - 2.0 Z$_{\odot}$, while models consistent with SN 2013aa${\dagger}$ and 2015F have metallcities in 0.0 Z$_{\odot}$ - 1.0 Z$_{\odot}$.

We note that all the near-$M_{\rm Ch}$ simulation models that are consistent with the observational events have metallicities in the range of 0.0 Z$_{\odot}$ to 3.0 Z$_{\odot}$. This is in contrast with the work presented by \cite{martinez2017observational}, who find the metallicity of the WD stellar progenitors of SNR 3C 397 and G337.2 to be 5.4 Z$_{\odot}$. These very high metallicity values might be because the models to which they compared their observations, only explored a range of metallicities values, and the central densities of the WD stellar progenitors were not explored (2 - 3 $\times 10^9$ g cm$^{-3}$). A higher central density stellar progenitor WD could suggest a high degree of neutronization for SNR 3C 397 as suggested by \cite{daveetal17}. We note that the metallicity values found by \cite{martinez2017observational} for SN remnants Tycho, N103B, and Kepler were below 2Z$_{\odot}$, and some of the models consistent with the events have similar metallicities. These metallicity values between $\sim$ 1.5Z$_{\odot}$ to $\sim$ 3Z$_{\odot}$ are on the higher end of the Galactic metallicity sprectrum \citep{rix2013milky}.

\subsection {Near-$M_{\rm{Ch}}$ WD Progenitors: Central Densities}

From the $\chi^2$ analysis, we find that the event SN 2012cg*, which has a high 57/56 ratio ($>$ 0.05), points to a near-$M_{\rm{Ch}}$ WD progenitor, with a very high central density. The models which are consistent with SN 2012cg* have central densities $\approx 6 \times 10^9$ g cm$^{-3}$. This result is in contrast with the work presented by \cite{iwamoto1999nucleosynthesis}, which argued that the central densities of an average progenitor WD in the near-$M_{\rm Ch}$ channel should be $\lessapprox 2 \times 10^9$ g cm$^{-3}$, so as to avoid the large ratios of $^{54}$Cr/$^{56}$Fe and $^{50}$Ti/$^{56}$Fe. Moreover, we note that the very high fraction of 57/56 and 55/56 for the consistent models of the event SN 2012cg$^*$ suggests a near-$M_{\rm{Ch}}$ WD progenitor similar to that of SNR 3C 397. Moreover, we find that SN 2015F is consistent with low central density models of the order $\approx 10^6$ g cm$^{-3}$ where the primary WD have masses in the range from 0.6 M$_{\odot}$ - 0.7 M$_{\odot}$.

\section{Conclusions}
\label{sec:LateTimeConclusion}
We have looked at a set of five nearby SN Ia events (SN 2011fe, SN 2012cg, SN 2013aa, SN 2014J, 2015F) for which the ratio of $^{57}$Ni/$^{56}$Ni was determined from their late-time light curves. We perform a $\chi^2$ test to constraint the parameters of these events with an extensive suite of simulation models in the near-$M\rm _{ch}$ and sub-$M\rm_{ch}$ progenitors.
We find the low and solar metallicity sub-$M_{\rm Ch}$ collision models of \citet{papish2016supernovae} and \citet{ZenatiPerets20} to be consistent with the event SN 2015F. Moreover, we find the high central density DDT and GCD models of \citet{daveetal17} to be consistent with the measurements made by \citet{graur2018observations} for SN 2012cg. Models from \citet{papish2016supernovae}, \citet{shen2018sub}, \citet{ZenatiPerets20} and \citet{leung2020explosive} in the double detonation and pure detonation of a sub-$M_{\rm Ch}$ were found to be consistent with the very late-time measurements by \citet{tucker2021whisper} for the event SN 2011fe.

The measurements of SN 2011fe by \citet{graur2016late} and \citet{shappeeetal16} can be explained by both the near-$M_{\rm Ch}$ and sub-$M_{\rm Ch}$ progenitors. For SN 2014J, no model agrees with the measurement made by \citet{graur2018late}, but both near-$M_{\rm Ch}$ and sub-$M_{\rm Ch}$ progenitor models are consistent with the measurement made by \citet{li2019observations}. Finally, both the single and double degenerate progenitor models are consistent with the measurements of SN 2012cg made by \citet{graur2016late} and of SN 2013aa made by \citet{jacobson2018constraining}.

We note that all simulation models have used different nucleosynthesis networks \citep{timmes99,travaglio2004nucleosynthesis,paxton2010modules} to calculate the yields of the 55, 56, and 57 chain nuclides, with differing nuclear reaction network sizes. The determination of formal uncertainties is a challenging task even for simpler 1D spherically-symmetric stellar evolutionary models \citep{fields2016properties}. As such, the formal uncertainties for multidimensional explosion models are generally left unreported. Nonetheless, because widely varying methods produce similar outcomes for broadly-similar initial conditions of sub-Mch versus near-Mch progenitors, the basic conclusions arrived at here are likely robust.

In a recent paper, \cite{polin2019nebular} calculated the synthetic spectra of the nebular phases for sub-$M_{ch}$ double detonation models and found it difficult to explain the excessive flux corresponding to $[\rm Ca\,\textsc{ii}]$ emissions in the models as compared to the spectra of SN 2011fe, a normal SN Ia. However, they found a good fit for SN 1999by, which is a 1991bg-like sub-luminous SN Ia. These results are consistent with our, where 2015F, which is a sub-luminous SN Ia, is consistent with a sub-$M_{\rm{Ch}}$ progenitor.

We conclude that normal SNe Ia cannot be explained by a single progenitor or explosion channel and both the single degenerate and the double degenerate channels are at play here. This result is consistent with the work presented by \citet{flors2020sub}, where the authors looked at optical spectra of 58 SNe Ia and concluded that 85\% of them were in an agreement with a sub-$M_{\rm Ch}$ progenitor while 15\% were in agreement with a near-$M_{\rm Ch}$ progenitor. Our results are also in alignment with the work of \citet{eitner2020observational}, where the authors compared the Galactic abundance of Mn/Fe and galactic chemical evolution models from different sources of SNe Ia and SNe II and found a good fit for the observations to the models with $\sim$ 75\% of the SNe Ia coming from the sub-$M_{\rm Ch}$ progenitors and $\sim$ 25\% from the near-$M_{\rm Ch}$ progenitors. On the other hand, \citet{kobayashi2020new} calculated chemical evolution models, and from the evolutionary trends of the elemental abundance ratios among Mn, Ni, Cr, and Fe, they concluded that the contribution of sub-$M_{\rm Ch}$ progenitors is less than 25\% in the solar neighborhood, while it is higher in dSph. Moreover, \citet{seitenzahl2019optical} used supernova remnant tomography to put constraints on three young SNe Ia remnants (SNR 0519-69.0, SNR 0509-67.5, and N103B) and found SNR 0519-69.0 to be consistent with a near-$M_{\rm Ch}$ progenitor and SNR 0509-67.5 to be consistent with a sub-$M_{\rm Ch}$ progenitor.

Finally, to put tighter constraints on the nature of the progenitors, future observations of SNe Ia metallicity can be used as prior information to the $\chi^{2}$ analysis. Moreover, as our knowledge of detonation physics improves, SNe Ia simulation models with physics-informed detonation initiation will help reduce the dimensionality of the explosion model parameter space.

\section*{Acknowledgements}

VT was supported by NASA through \textit{HST}-GO-15693. OG was supported, in part, by an NSF Astronomy and Astrophysics Fellowship under award AST-1602595. RTF thanks the NASA ATP program for support under award  80NSSC18K1013. RTF also thanks the Institute for Theory and Computation at the Center for Astrophysics \textbar\ Harvard \& Smithsonian, and the Kavli Institute for Theoretical Physics, supported in part by the national Science Foundation under grant NSF PHY11-25915, for visiting support during which this work was undertaken. SCL acknowledges support from NASA grants HST-AR-15021.001-A and 80NSSC18K1017.
 KN is supported by the World Premier International Research Center
Initiative (WPI Initiative), MEXT, Japan, and JSPS KAKENHI Grant
Numbers JP17K05382 and JP20K04024, and JP21H04499. KJS is supported by NASA through the Astrophysics Theory Program (NNX17AG28G).

\section*{Data Availability}

The data behind Figure \ref{fig:57by56plot} and Figure \ref{fig:55by56plot} is available here: https://doi.org/10.5281/zenodo.5797233 .



\bibliographystyle{mnras}
\bibliography{mnras_template} 








\bsp	
\label{lastpage}
\end{document}